# Investigating the Relationship between Freeway Rear-end Crash Rates and Macroscopically Modelled Reaction Time

Ishtiak Ahmed[a]*, Billy M. Williams[b], M. Shoaib Samandar[c], Gyounghoon Chun[a]

*Corresponding Author's Email: iahmed2@ncsu.edu*

[a]*Department of Civil, Construction, and Environmental Engineering, North Carolina State University, 909 Capability Drive, Raleigh, NC 27606.*

[b]*Director, Institute for Transportation Research and Education (ITRE), North Carolina State University, 909 Capability Drive, Raleigh, NC 27606*

[c]*Research Associate, Institute for Transportation Research and Education (ITRE), North Carolina State University, 909 Capability Drive, Raleigh, NC 27606*

Word Count: 6,305 words + 2 tables + 6 figures

*Submitted [August 22, 2020]*

# Investigating the Relationship between Freeway Rear-end Crash Rates and Macroscopically Modelled Reaction Time

## ABSTRACT

This study explores the hypothesis that an analytically derived estimate of the required driver reaction time for asymptotic stability, based on the macroscopic Gazis, Herman, and Rothery (GHR) model, can serve as an effective indicator of the impact of traffic oscillations on rear-end crashes. If separate GHR models are fit discontinuously for the uncongested and congested regimes, the local drop in required reaction time between the two regimes can also be estimated. This study evaluates the relationship between freeway rear-end crash rates and this drop in driver reaction time.

Traffic data from 28 sensors collected over one year were used to calibrate the two-regime GHR model. Rear-end crash rates for the segments surrounding the sensor locations are estimated using archived crash data over four years. The rear-end crash rates exhibited a strong positive correlation with the reaction time drop at the density-breakpoint of the congested regime. A linear form model provided the best fit in terms of R-square, standard error, and homoscedasticity. These results motivate follow-on research to incorporate macroscopically derived reaction time in road-safety planning. More generally, the study demonstrates a useful application of a discontinuous macroscopic traffic model.

## INTRODUCTION

Understanding traffic characteristics that may lead to rear-end crashes is not only important for improving safety, but also important to alleviate non-recurrent congestion on the roads. Past studies showed that rear-end crashes are strongly associated with traffic instability and oscillations (Tanaka, Ranjitkar, and Nakatsuji 2008, 198-207; Touran, Brackstone, and McDonald 1999, 567-578) . Due to traffic oscillations, drivers often fail to react in time and collide with the vehicle in front. To identify rear-end crash-prone locations, the state of the practice is documented in the Highway Safety Manual (American Association of State Highway and Transportation Officials 2010) . However, highway safety engineering is founded on statistical analysis of crash occurrences. Therefore, the identification of rear-end crash-prone locations would be improved if the crash risk could be predicted using non-crash data, for instance, the macroscopic traffic characteristics of a site.

The steady state macroscopic flow model equivalent to the so-called Gazis-Herman-Rothery (GHR) (Gazis, Herman, and Rothery 1961, 545-567) car-following model provides a rational means to estimate the reaction time necessary for asymptotic stability (May 1965, 290-303). This model can be fitted to observed macroscopic traffic data and subsequently, to analytically estimate the driver reaction time necessary for asymptotic stability. Here, it is hypothesized that this analytically estimated driver reaction time required for asymptotic stability can serve as an effective indicator of the impact of traffic oscillations on rear-end crashes.

This study investigates the relationship between freeway rear-end crash rates and macroscopically derived reaction time required for asymptotic stability. Since rear-end crashes tend to occur more during traffic state transitions, it specifically focuses on the change in required reaction time when traffic state shifts from the uncongested to the congested regime. To investigate this change, a discontinuous form of macroscopic GHR model is calibrated to traffic sensor data. The relationship between long-term rear-end crash rates of a freeway segment and change in required reaction time at that location is investigated using regression analysis.

This paper is organized as follows. A review of past studies on the topic is presented in the next section. The following section presents the mechanism of a discontinuous form of GHR macroscopic model development, driver reaction time estimation, and rear-end crash rate estimation from the selected freeway sites. Results from the fitted models, rear-end crash analysis, and statistical regression analysis are discussed in the following section. Finally, the summary of the results and limitations of the study is presented in the last section.

**LITERATURE REVIEW**

May (1965, 290-303) defined an unstable driver behaviour as when a following driver responds slowly to the change in the speed of the leading vehicle, but when responds, exerts large acceleration or deceleration rates. Therefore, the stability of car-following behaviour of two vehicles, or in a broader term, of a traffic stream is a function of the driver reaction time and sensitivity. The General Motor researchers developed an analytical formulation of the sensitivity term and showed that the product of this sensitivity and reaction time dictates the stability of car-following behaviour both on a local and asymptotic scale. Details on the mathematical condition of asymptotic stability are discussed later in this paper. In recent decades, several studies investigated the relationships among these critical car-following parameters, namely, reaction time, sensitivity, headway (Kim and Zhang 2011, 52-61; Zielke, Bertini, and Treiber 2008, 57-67; Ahn, Laval, and Cassidy 2010, 1-8; Xu and Laval 2019, 610-619) . Kim and Zhang investigated the relationship among these parameters based on the GHR and other simpler car-following models using car-following behaviour data extracted from the NGSIM (USDOT 2006) dataset. Another study (Zielke, Bertini, and Treiber 2008)  conducted a comparative investigation of traffic oscillation in terms of the amplitude, propagation velocity, and frequency of shockwave across three countries. Ahn et al. (2010, 1-8)  analytically proved with a triangular-shaped fundamental diagram that the amplitude of an oscillation through a queued traffic stream dampens as it passes an on-ramp and magnifies as it passes an off-ramp.

In addition to car-following behaviour and traffic stability analysis, a few studies extended their focus on these parameters' effects on crash risk  (Tanaka, Ranjitkar, and Nakatsuji 2008, 198-207; Misener et al. 2000, 29-38; Chatterjee and Davis 2016, 110-118; Davis and Swenson 2006, 728-736) . Tanaka et al. (2008) used a microscopic dataset of ten trajectories to investigate the effect of reaction time and sensitivity on asymptotic stability using different safety indicators. Although several safety indicators showed that traffic oscillation propagates downstream when the product of sensitivity and reaction time exceeds the threshold proposed by Chandler et al. (Chandler, Herman, and Montroll 1958, 165-184) , other safety indicators yielded an inconclusive result. Nonetheless, it successfully demonstrated the effect of reaction time on traffic stability. To explore the causal relationship between crash occurrence and following headway along with reaction time, Davis and Swenson used video recorded microscopic traffic data and the kinematic theory developed by Brill (1972, 343-353)  and simulated three real-world rear-end crashes. It revealed that had the colliding vehicles or a few vehicles further downstream in the sequence maintained a higher following headway than their reaction time, the collisions would probably have been avoided. Chatterjee and Davis extended the analytical formulation of crash occurrence using car-following theory by Brill for a series of vehicles in a platoon. It perceived the stopping distance for the braking of the first car in a platoon as a shared resource and this resource is either consumed or contributed by the following vehicles. If the sum of the consumption of this resource by the following vehicles exceeds a threshold, rear-end crash occurs. This theory was verified by observing 41 shockwaves in which 15 rear-end crashes or swirving events to avoid crashes were occurred. To develop a probabilistic model for rear-end crash occurrence using trajectory data, Oh and Kim (2010, 1888-1893) used two probability measures, namely probability of changing lanes and probability of the following vehicle hitting the leading one for a given "Time to Collision". However, the accuracy of the proposed approach was not tested against field crash data. While the availability of microscopic information on driver reaction time level is increasing with the increasing research using instrumented vehicles (Tanvir, Chase, and Roupahil 2019, 1-14) , microscopic traffic data like acceleration, following headway, and instantaneous reaction time is still difficult to obtain by transport agencies. Therefore, the practical application of the studies discussed above is limited since these demand several micro-level inputs like acceleration, following headway, and reaction time.

Several studies attempted to predict the rear-end crash risk of a roadway using macroscopic traffic data (Lord, Manar, and Vizioli 2005, 185-199; Abdel-Aty et al. 2004, 88-95). Among these, Lord et al. used traffic density and volume-to-capacity ratio as explanatory variables to predict freeway crashes. Abdel-Aty et al. showed that a combination of high coefficient of variation of speed and high occupancy at a downstream segment is a potential crash pre-cursor in the upstream segment of a roadway. Pande and Abdel-Aty (2006, 31-40) divided rear-end crashes into two groups based on whether they occur before or during congestion. Average and coefficient of variation of speed, average occupancy, and presence of a downstream ramp were the statistically significant variables to predict rear-end crashes. However, the discrepancies in the findings of these studies indicate that the findings are mostly site-specific.

The above survey of literature shows that while many studies analyzed the car-following model parameters related to driver reaction time, only a few focused on its relation to rear-end crash rates. These few studies used microscopic trajectory level data that are difficult to obtain on a network-level. Although several studies investigated the crash precursor potential of various macroscopic traffic characteristics, those mostly focused on a single site and the findings are mostly site-specific.

**METHODOLOGY**

In this section, first, the macroscopic model equivalent to the GHR car-following model is explained. Next, the derivation of driver reaction time required for asymptotic stability is described. Finally, the description of the study site and crash data collection and analysis method are presented.

***The Macroscopic Model Equivalent to the GHR Car-following Model***

The basic form of the macroscopic model equivalent to the fifth GHR car-following model *(4)* is:

$$q_{M,i} = k_i * u_f * \left( 1 - \left( \frac{k_i}{k_j} \right)^{l-1} \right)^{\frac{1}{1-m}} \tag{1}$$

Where, $i$=1,2,3 . . . Observation index; $q_{M,i}$= Model flow for observation $i$ (pc/hr/ln); $u_f$= Free flow speed (mph); $k$= Observed density (pc/mi/ln); $k_j$=Jam density (pc/mi/ln); $l$= Distance headway exponent; $m$= Speed difference exponent.

Variety of versions of this model form have been proposed by past studies by incorporating multiple regimes and discontinuity in the fundamental diagram. Since our target was to derive the drop in driver reaction time during the transition of two regimes, we adopted the so called inverse lambda shaped flow-density model form.

A few versions of the inverse lambda shaped flow-density form is found in the past studies (Edie 1961, 66-76; Ahmed, Williams, and Samandar 2018, 51-62) . Among these, a relatively recent study by Ahmed et al. proposed a two-regime fundamental diagram with an overlap of the regimes near the capacity. Moreover, this study also proposed an algorithm to filter the steady-state observations from side-fire radar sensors. A slightly modified version of that approach is adopted in this study.

The proposed modeling algorithm can be divided into three steps: i) data pre-processing to filter steady-state observations ii) initial fitting of a two-regime GHR Model iii) iteration of the second step with robust regression to remove remaining outliers. For the details of the first step, readers are suggested to review the previous research (Ahmed et al. 2018).

To fit the model shown in Equation 1 with a transition regime, it was assumed that the transition regime includes a mixture of observations from both uncongested and congested regimes. Empirical observations of traffic stream data also depict that the data points in the transition range follow either the uncongested or the congested regime's characteristics. Within the overlap of the uncongested and congested regime, each data point was proposed to be modelled by the regime model, either uncongested and congested, that results in the smallest absolute error. **Equation 2, 3, and 4** demonstrates the mechanism of the proposed model.

| Observed Density | Formula for model flow |
|---|---|

$$k_i \leq kb_{r=2} \qquad q_{M,i,r=1} = k_i * u_{f1} * \left(1 - \left(\frac{k_i}{k_{j1}}\right)^{l_1 - 1}\right)^{\frac{1}{1-m_1}} \qquad (2)$$

$$k_i \geq kb_{r=1} \qquad q_{M,i,r=2} = k_i * u_{f2} * \left(1 - \left(\frac{k_i}{k_{j2}}\right)^{l_2 - 1}\right)^{\frac{1}{1-m_2}} \qquad (3)$$

$$kb_{r=2} < k_i \qquad q_{M,i,r} = \begin{cases} q_{M,i,r=1} \; if \; |q_{M,i,r=1} - q_i| < |q_{M,i,r=2} - q_i| \\ q_{M,i,r=2} \; Otherwise \end{cases} \qquad (4)$$
$$< kb_{r=1}$$

Here, $r$ = Regime index. 1 represents uncongested and 2 represents congested regime.

$kb$ = Density breakpoint. Note that $kb_{r=2} < kb_{r=1}$.

In **Equation 2** through **4**, the two density breakpoints ($kb_{r=1}$ and $kb_{r=2}$) define the upper and lower limit of the transition regime, respectively.

The above algorithm fits an inverse lambda shaped flow-density model to the observed data. However, it could not capture some high-flow observations near the capacity of a segment. Moreover, the queue discharge flow rate according to the model was significantly lower than the expected range. To tackle these issues, two thresholds based on the Highway Capacity Manual's (HCM) basic freeway segment analysis method were incorporated in the algorithm (Transportation Research Board of the National Academies 2016). According to the first threshold, the slope at capacity of the flow-density model must be less than or equal to the HCM derived slope at capacity value. The second threshold implies that the queue discharge flow rate should be within 2% to 20% of the capacity as specified in the HCM. Application of these thresholds resulted in a well-fitted model with reasonable estimates of capacity and queue discharge flow rate. Note that a third threshold was applied in the past study to keep the jam density within a reasonable range. However, that threshold is skipped here as empirical observation showed that artificially capping the jam density value may result in a poor fit of the congested regime curve to the observed data.

To further remove outliers from the field observation, the so-called Robust Regression technique was used to fit the model and remove outliers based on the fitted model iteratively. From the initial model fit, the standard error for each data point is estimated. Data points with a standard error higher than a certain threshold are removed from the original dataset. Then, the model is fitted again with the updated dataset. The process is continued until the maximum standard error becomes lower than the threshold. The formula for estimating standard error (SE) in terms of flow is shown in **Equation 5**.

$$SE_{i,r} = (q_{M,i,r} - q_i)/Std_r \qquad (5)$$

Where, $Std_r$ = Standard deviation of flow error for regime $r$, and $q_i$ is the observed error.

The major issue with the selection of the threshold was that while measurement errors are present in both regimes, mixed-state observations are more prevalent in the congested regime than in the uncongested regime. To remove both measurement errors and mixed-state data, two different thresholds are applied to the two regimes. To remove the measurement errors from the uncongested regime which are symmetric in nature, a symmetric threshold of ±3.5 is applied to the uncongested regime standard error. Mixed state observations, on the other hand, are likely to be distributed asymmetrically as these tend to be on the left side of the flow-density curve. Hence, an asymmetric threshold of +2 is applied for all but the final step of the robust regression. In the final step of robust regression, a threshold of -3.5 is applied to exclude any remaining outliers from the congested regime on the right side of the flow-density curve.

***Estimating Required Driver Reaction Time for Different Traffic States***

The formula for estimating the required driver reaction time for asymptotic stability needs to be derived from the microscopic form of the fifth and final GHR car following model, which is expressed in **Equation 6**. Here, the acceleration of the *(n+1)th* vehicle in a traffic stream at time $(t + \Delta t)$ (termed as $x''_{n+1}(t + \Delta t)$) in response to the relative speed between the nth and *(n+1)th* vehicle at time t is expressed as the product of the sensitivity term and the relative speed between the two vehicles.

$$x''_{n+1}(t + \Delta t) \ = \alpha \frac{[x'_{n+1}(t + \Delta t)]^m}{[x_n(t) - x_{n+1}(t)]^l} * [x'_n(t) - x'_{n+1}(t)] \qquad (6)$$

Here, *n*= Position of a driver in a traffic stream. (*n*=0 is the most downstream driver).

$x_n$= Location of the nth driver with respect to a reference point.

$x'_n(t)$= Speed of the nth driver at time *t*

According to May (1965) , the parameter $\alpha$ can be expressed as shown in **Equation 7**.

$$\alpha = \frac{(l-1)u_f^{1-m}}{(1-m)k_j^{l-1}} \qquad (7)$$

Thus, the sensitivity factor is equivalent to what is shown in **Equation 8**

$$Sensitivity\ factor = \frac{(l-1)u_f^{1-m}}{(1-m)k_j^{l-1}} * \frac{[x'_{n+1}(t+\Delta t)]^m}{[x_n(t)-x_{n+1}(t)]^l} \tag{8}$$

For steady state observations, individual vehicle speed represents the speed of the traffic stream and the spacing between two successive vehicles represents the inverse of the traffic density. Thus, **Equation 9** can be written as

$$Sensitivity\ factor = \frac{(l-1)u_f^{1-m}}{(1-m)k_j^{l-1}} * \frac{u^m}{\left(\frac{1}{k}\right)^l} \tag{9}$$

According to May (1965) , for a traffic stream to be asymptotically stable, the product of the reaction time and sensitivity must be less than or equal to 0.5. Hence, the expression for the minimum reaction time required for asymptotic stability can be derived as:

$$t_i = \frac{(1-m)k_j^{l-1}}{2(l-1)u_f^{1-m}} * \frac{\left(\frac{1}{k_i}\right)^l}{u^m} \tag{10}$$

**Equation 10** gives the formula to estimate the driver reaction time required for stability for each observation of flow, speed, and density.

*Drop in Required Reaction Time between Two Regimes*

Fitting the proposed discontinuous flow-density model as shown earlier enabled the research team to investigate the drop in driver reaction time when the traffic state moves from uncongested to the congested regime. If the reaction time described in **Eq. (10)** is estimated for a series of density values, the following curves are obtained.

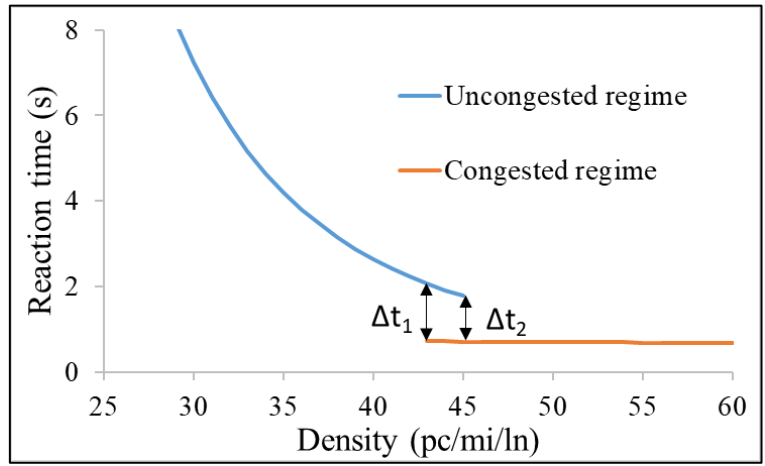

Figure 1 A typical required reaction time vs. density plot

Upon scrutinizing the same plot of reaction time vs. density for different locations, it was found that the reaction time for the congested regime does not vary significantly. Moreover, the reaction time for a very low density value (such as less than 20) may not have any significance as traffic stream barely follows the car-following model. On the other hand, if the transition regime is focused here (bounded by the two vertical arrows in **Figure 1**), there are two reaction times for each density point within this regime. As the traffic state transfers from the uncongested to the congested regime, the required driver reaction time also drops to get adapted to the change in traffic state. The research team hypothesizes that higher the drop in reaction time, higher the risk of a rear-end crash to occur.

The two drops in reaction time shown in **Figure 1** are of particular interest here. These two drops termed as $\Delta t_1$ and $\Delta t_2$ are the drops in reaction time at the beginning and the end of the transition regime, respectively. $\Delta t_1$ is the largest drop in reaction time within the overlap. It is the value by which the minimum driver reaction time needs to be changed to maintain asymptotic stability when traffic state transfers from the uncongested to the congested regime at a density of $kb_2$. On the other hand, $\Delta t_2$ is the change in minimum driver reaction time when traffic state transfers the regime upon reaching the capacity and at a density of $kb_1$. The exploration of correlation between rear-end crash rates and drop in driver reaction time in this study revolves around these two extreme drops in driver reaction time.

### Data Description

*Traffic Data from Sensors*

In this study, the discontinuous macroscopic model described above is proposed to be fitted with field data collected from side-fire radar sensors located on different locations of the freeway system of the Triangle Region of North Carolina. Flow, speed, and lane occupancy data from 28 directional sensors are collected for the calendar year of 2013 in a time resolution of 5 minutes. These sensors are located on three interstates namely I-40, I-440, and I-540. Since this study primarily hinges on driver car-following model, only basic freeway segments (see HCM for definition) are selected to ensure that the lane changing activities are minimum near the sensors. **Figure 2** shows the location of these sensors in the study area. Past studies showed that several recurring bottlenecks exist in the proximity of some of these locations (Ahmed, Rouphail, and Tanvir 2018, 235-246) . The numbers show the tag of each station and the alphabets attached to them indicate the travel direction.

Prior to fitting the model with these data, a filtering algorithm was applied to remove mixed-state and inconsistent observations as much as possible. Details of the filtering algorithm are presented in a past study (Ahmed, Williams, and Samandar 2018, 51-62).

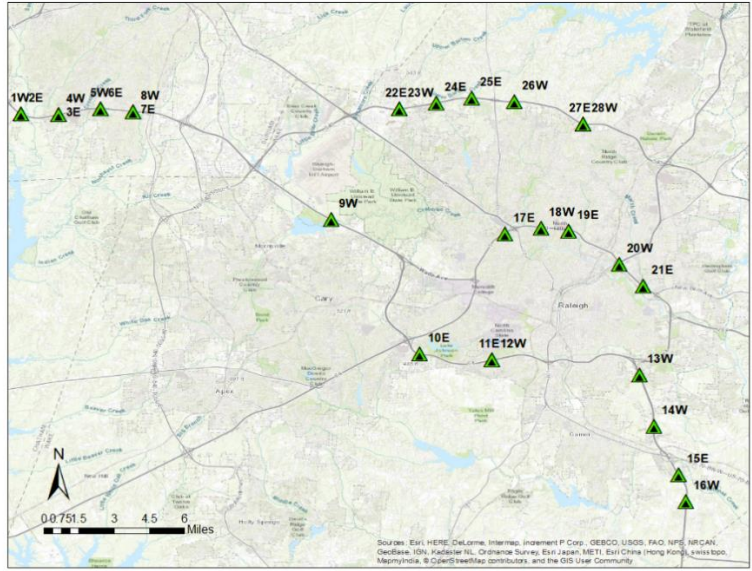

Figure 2 Location of the side-fire radar sensors in the study area

*Crash Data*

To estimate the rear-end crash rate associated with each sensor, the most critical task was to select an appropriate freeway segment surrounding the sensor location. As explained in the Highway Safety Manual, the crash rate at a segment can be attributed to several geometric features including the vertical and horizontal curvature, lane width, number of lanes, and presence of ramps. Here, each segment was selected in such a way that it is away from any ramp, does not have any tight curve, and the number of lanes and lane width is consistent throughout the segment.

Four years of police-reported crash data from the selected segments are collected for the period from 2011 to 2014. A tool called "Traffic Engineering Accident Analysis System (TEAAS)" (North Carolina Department of Transportation) was used to extract the police reports. Each crash report was carefully investigated to decide if that crash occurred within the corresponding segment. The average AADT of the selected years is obtained from the archive of North Carolina Department of Transportation. Finally, the following equation was used to estimate the rear-end crash rate (crashes per 100 million VMT) for each segment.

$$Crash\ rate = 100,000,000 * C/(365 * N * \overline{AADT} * 0.5 * L) \qquad (11)$$

Where $N$ = number of years over which crash data were collected ($N = 4$)

$C$ = total frequency of crashes in $N$ years.

$\overline{AADT}$ = average AADT over $N$ years.

$L$ = segment length in miles.

After estimating the crash rate for each segment, its correlation with the two drops in required reaction time ($\Delta t_1$ and $\Delta t_2$) are investigated using statistical modeling through regression analysis.

## ANALYSIS AND RESULTS

This section is divided into three major parts. In the first part, results from fitting the proposed two-regime traffic flow model is presented. The fitted parameter values and estimated drops in required reaction time are also discussed here. In the second part, results from the rear-end crash rate analysis are described. Finally, the relationship between driver reaction time drop and rear-end crash rates are assessed.

### *Fitted Two-Regime Traffic Flow Models*

To fit the traffic stream model by imposing the constraints described earlier, a nonlinear optimization tool available in MATLab was used. The fundamental diagrams for the sensor 18W is shown **Figure 3**. The parameter values used to plot these diagrams are the fitted parameters obtained from the convergence of the proposed robust regression algorithm. **Table 1** presents the estimated parameter values, their standard deviations, and the resulting required reaction time drops ($\Delta t_1, \Delta t_2$). Standard deviation for such nonlinear optimization models is calculated using a method described in a past study (Smith 2013).

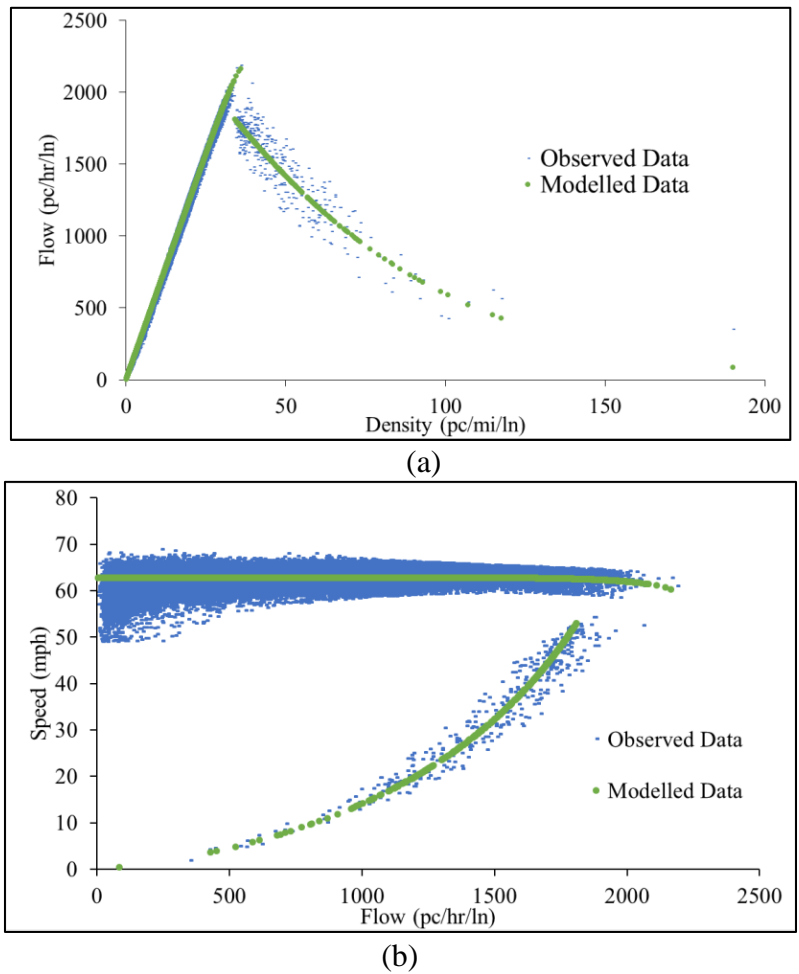

(a)

(b)

Figure 3 Fundamental diagrams for sensor 18W (a) Flow vs. Density, (b) Speed vs. Flow

TABLE 1 Fitted Parameter Values for Different Sensors and the Resulting Reaction Time Drops

| ID | Uncongested Regime | | | | | Congested Regime | | | | | $\Delta t_1$ | $\Delta t_2$ |
|---|---|---|---|---|---|---|---|---|---|---|---|---|
| | $u_f$ | $k_b$ | $k_j$ | $l$ | $m$ | $u_f$ | $k_b$ | $k_j$ | $l$ | $m$ | | |
| 1W | 65 (5.9E-3) | 40 | 88.9 (9.1E-2) | 4.418 (2.8E-3) | 0.609 (1.4E-3) | 4.7E5 (4.7E2) | 35 | 466 (3.5E-1) | 1.022 (7.1E-6) | 0.684 (3.3E-5) | 1.51 | 0.68 |
| 2E | 62 (8.0E-3) | 39 | 81.5 (4.6E-1) | 10.37 (5.7E-2) | 0.985 (8.0E-4) | 4.5E5 (4.9E2) | 35 | 467 (3.3E-1) | 1.022 (7.9E-6) | 0.677 (3.6E-5) | 2.82 | 0.8 |
| 3E | 56.4 (6.9E-3) | 42 | 129.8 (5.3E+0) | 9.333 (2.5E-1) | 0.99 (3.4E-3) | 1.2E7 (9.2E3) | 38 | 337 (1.7E-1) | 1.003 (9.3E-7) | 0.594 (2.4E-5) | 2.63 | 0.92 |
| 4W | 59.2 (6.0E-3) | 39 | 237.5 (3.1E+1) | 8.215 (4.5E-1) | 0.99 (9.5E-2) | 4.4E5 (2.9E2) | 35 | 641 (7.3E-1) | 1.04 (1.8E-5) | 0.757 (4.5E-5) | 2.67 | 0.70 |
| 5W | 61.7 (5.0E-3) | 44 | 209.6 (2.2E-1) | 5.47 (2.5E-3) | 0.993 (3.3E-5) | 1E5 (1.7E2) | 39 | 2133 (2.1E+0) | 1.134 (3.5E-5) | 0.884 (2.5E-5) | 1.8 | 0.86 |
| 6E | 57.7 (1.1E-2) | 50 | 614.2 (9.4E-1) | 3.964 (1.6E-3) | 0.997 (1.4E-5) | 3.3E5 (3.2E2) | 45 | 1077 (6.2E-1) | 1.062 (1.3E-5) | 0.808 (2.0E-5) | 1.51 | 0.83 |
| 7E | 63.3 (1.0E-2) | 39 | 76.7 (5.9E-1) | 11.016 (9.0E-2) | 0.981 (1.5E-3) | 5.7E6 (9.5E3) | 36 | 918 (9.9E-1) | 1.025 (9.5E-6) | 0.781 (3.0E-5) | 2.25 | 0.55 |
| 8W | 61.7 (9.5E-3) | 42 | 146 (3.2E-1) | 6.716 (8.5E-3) | 0.993 (8.7E-5) | 6.2E6 (1.3E4) | 38 | 229 (3.4E-1) | 1.002 (2.0E-6) | 0.538 (7.9E-5) | 1.56 | 0.69 |
| 9W | 65.3 (6.5E-3) | 48 | 572 (1.8E+1) | 4.862 (4.2E-2) | 0.99 (1.2E-3) | 1.4E6 (1.3E3) | 40 | 1800 (8.1E-1) | 1.11 (1.4E-5) | 0.894 (9.7E-6) | 2.19 | 0.7 |
| 10E | 62.9 (1.2E-2) | 42 | 399.9 (5.7E-1) | 2.973 (1.0E-3) | 0.955 (1.3E-4) | 3.4E5 (9.4E2) | 38 | 1307 (1.8E+0) | 1.116 (4.8E-5) | 0.877 (3.7E-5) | 1.67 | 1.15 |
| 11E | 63.6 (1.3E-2) | 37 | 366.4 (4.9E-1) | 4.149 (1.6E-3) | 0.996 (1.7E-5) | 3.4E5 (5.9E2) | 37 | 505 (6.8E-1) | 1.029 (1.6E-5) | 0.702 (5.8E-5) | 0.89 | 0.89 |
| 12E | 65.1 (4.8E-3) | 40 | 451.7 (5.1E-1) | 4.352 (1.3E-3) | 0.998 (7.6E-6) | 3.4E5 (8.9E2) | 36 | 447 (4.5E-1) | 1.119 (5.5E-5) | 0.846 (4.2E-5) | 2.14 | 1.63 |
| 13W | 61.9 (6.6E-3) | 44.6 | 342.1 (6.0E+0) | 5.609 (3.6E-2) | 0.99 (8.1E-4) | 1E5 (1.3E2) | 42 | 1043 (7.3E-1) | 1.119 (2.8E-5) | 0.849 (2.4E-5) | 1.27 | 0.88 |
| 14W | 61.4 (8.1E-3) | 45 | 729.5 (6.6E-1) | 3.395 (7.1E-4) | 0.994 (1.3E-5) | 3.4E5 (2.3E2) | 45 | 340 (1.8E-1) | 1.007 (2.3E-6) | 0.519 (3.6E-5) | 0.74 | 0.74 |
| 15E | 62.4 (9.1E-3) | 40 | 1099.7 (5.9E+0) | 1.973 (1.4E-3) | 0.205 (4.2E-3) | 1E5 (5.3E1) | 36 | 1812 (6.9E-1) | 1.091 (9.3E-6) | 0.839 (1.1E-5) | 2.17 | 0.95 |
| 16W | 67.1 (1.6E-2) | 42 | 183 (2.8E-1) | 3.056 (1.7E-3) | 0.792 (6.7E-4) | 2.9E6 (5.9E3) | 36 | 4207 (5.3E+0) | 1.066 (4.7E-7) | 0.879 (2.2E-5) | 1.33 | 0.96 |
| 17E | 67.8 (6.1E-3) | 40 | 196 (1.0E+0) | 6.09 (2.6E-4) | 0.99 (2.6E-4) | 1E5 (1.3E2) | 35 | 1808 (1.1E+0) | 1.174 (2.9E-5) | 0.907 (1.5E-5) | 2.01 | 0.84 |
| 18W | 62.8 (4.8E-3) | 37 | 69.4 (3.1E-1) | 11.207 (5.6E-2) | 0.971 (1.3E-3) | 5.4E5 (3.5E2) | 33 | 345 (1.3E-1) | 1.035 (6.6E-6) | 0.722 (1.9E-5) | 2.95 | 0.78 |
| 19E | 58.8 (4.9E-3) | 44 | 246 (1.1E+0) | 5.546 (9.7E-3) | 0.99 (2.0E-4) | 1E5 (2.1E2) | 39 | 609 (7.0E-1) | 1.091 (4.3E-5) | 0.802 (5.2E-5) | 2.82 | 1.47 |
| 20W | 63.3 (6.8E-3) | 40 | 606.2 (3.3E+0) | 3.846 (5.0E-3) | 0.99 (1.5E-4) | 4.7E5 (2.7E2) | 36 | 510 (2.0E-1) | 1.037 (5.8E-6) | 0.742 (1.6E-5) | 2.46 | 1.68 |
| 21E | 63.3 (3.9E-3) | 38 | 531.1 (9.8E-1) | 5.26 (2.7E-3) | 1 (7.8E-7) | 3.4E5 (7.6E2) | 37 | 269 (3.7E-1) | 1.023 (1.8E-5) | 0.652 (8.3E-5) | 0.98 | 0.98 |
| 22E | 66 (6.3E-3) | 39 | 79.7 (8.2E-2) | 5.83 (5.2E-3) | 0.752 (1.2E-3) | 4.6E5 (3.6E2) | 36 | 417 (2.2E-1) | 1.028 (6.8E-6) | 0.702 (2.5E-5) | 1.22 | 0.73 |
| 23W | 69.9 (6.3E-3) | 42 | 497.6 (3.8E-1) | 4.559 (1.0E-3) | 0.999 (2.7E-6) | 3.4E5 (2.4E2) | 39 | 852 (3.7E-1) | 1.074 (1.1E-5) | 0.817 (1.5E-5) | 1.2 | 0.75 |
| 24E | 63.8 (8.4E-3) | 44 | 427.7 (3.3E+0) | 4.512 (1.0E-2) | 0.99 (2.7E-4) | 4.3E5 (5.7E2) | 38 | 716 (5.4E-1) | 1.062 (1.8E-5) | 0.802 (2.8E-5) | 1.8 | 0.70 |
| 25E | 61.8 (9.8E-3) | 42 | 289.3 (5.4E+1) | 6.95 (5.2E-1) | 0.99 (1.1E-2) | 1E5 (1.1E2) | 39 | 1045 (7.1E-1) | 1.112 (2.4E-5) | 0.845 (2.3E-5) | 1.47 | 0.90 |

| ID | Uncongested Regime | | | | | Congested Regime | | | | | $\Delta t_1$ | $\Delta t_2$ |
|---|---|---|---|---|---|---|---|---|---|---|---|---|
| | $u_f$ | $k_b$ | $k_j$ | $l$ | $m$ | $u_f$ | $k_b$ | $k_j$ | $l$ | $m$ | | |
| 26W | 62.7 (8.7E-3) | 46 | 477.5 (9.3E-1) | 3.718 (2.0E-3) | 0.989 (5.8E-5) | 3.4E5 (1.5E2) | 43 | 555 (1.8E-1) | 1.024 (3.6E-6) | 0.679 (1.6E-5) | 1.35 | 1.07 |
| 27E | 67.7 (1.2E-2) | 41 | 473.1 (7.7E+0) | 4.403 (1.9E-2) | 0.99 (5.5E-4) | 1.6E6 (1.5E3) | 36 | 1426 (8.6E-1) | 1.055 (9.5E-6) | 0.834 (5.0E-6) | 1.34 | 0.68 |
| 28W | 68.1 (8.2E-3) | 41 | 447.7 (3.4E+0) | 4.188 (8.6E-3) | 0.99 (2.4E-4) | 6.8E6 (4.4E3) | 36 | 1289 (5.0E-1) | 1.043 (5.0E-6) | 0.833 (9.0E-6) | 1.72 | 0.81 |

The fundamental diagrams shown in **Figure 3** illustrate that the fitted models reasonably follow through the steady-state observations. Here, the value of the parameters that have physical interpretation needs to be discussed. The distance headway ($l$) and speed exponent ($m$) for both regimes, free flow speed for the congested regime, and jam density for the uncongested regime do not have any physical interpretations. The free flow speed of the uncongested regime varies from about 56 to 70 mph across these sites. The jam density of the congested regime varies over a very wide and somewhat unrealistic range of 229 to 4207 pc/mi/ln. This is due to the fact that observations near jam density condition is very scarce. Hence, the congested regime curve extrapolates to a very high density value when flow=0 in that regime. Unlike the previous research (Ahmed, Williams, and Samandar 2018, 51-62) , we are not artificially capping the jam density value which caused the congested regime to have a very poor fit to the observed data for some sensors.

The two density breakpoint values ($kb_{r=1}$ and $kb_{r=2}$) are of particular interests in this study since these two constitute the overlap and consequently, the required reaction time drops. Here, $kb_{r=1}$ ranges from 37 to 50 pc/mi/ln. However, in most sensors, it is less than the density at capacity (45 pc/mi/ln) for basic freeway segment specified by HCM. This wide-range variation of density breakpoint underscores that the national average value provided by HCM needs to be calibrated with field data if high fidelity analysis is desired.

The difference between the two density break points represents the overlap range which appears to be a unique characteristic for these sensors. The fitted overlap in density varies from 0 to 8 pc/mi/ln. The value for the first and second drop in driver reaction time vary from 0.74 to 2.95 and from 0.55 to 1.68, respectively, indicating that the sites have significant variations in the estimated drops in required reaction time.

The standard deviations of the estimates for the parameters to which the model is sensitive are of small magnitude, as shown within parenthesis in Table 1. This reveals that most of these estimates are significant and the error in the estimates of $\Delta t_1$ and $\Delta t_1$ should be within an acceptable range. It should be noted that the standard deviations for the two breakpoints ($k_b$) cannot be estimated because these are only classifiers of the regimes. Standard deviation for the congested regime free flow speed is very high because the fitted models are insensitive to this parameter. For the same reason, the standard deviation of the uncongested regime jam density for a few sensors are high (e.g., for 4W and 25E).

### Analysis of Crash Data

**Figure 4** shows the estimated rear-end crash rates (per 100 million VMT) for all the 28 sensors. The segment length, which ranges from about 0.15 to 2 miles, is showed in this figure by color-coding the bars. The figure shows that the selected sites have significant variability in terms of crash frequency since the crash rate and segment length appear to be uncorrelated. The crash rate spans over a wide range of about 10 to 65 rear-end crashes per 100 million VMT.

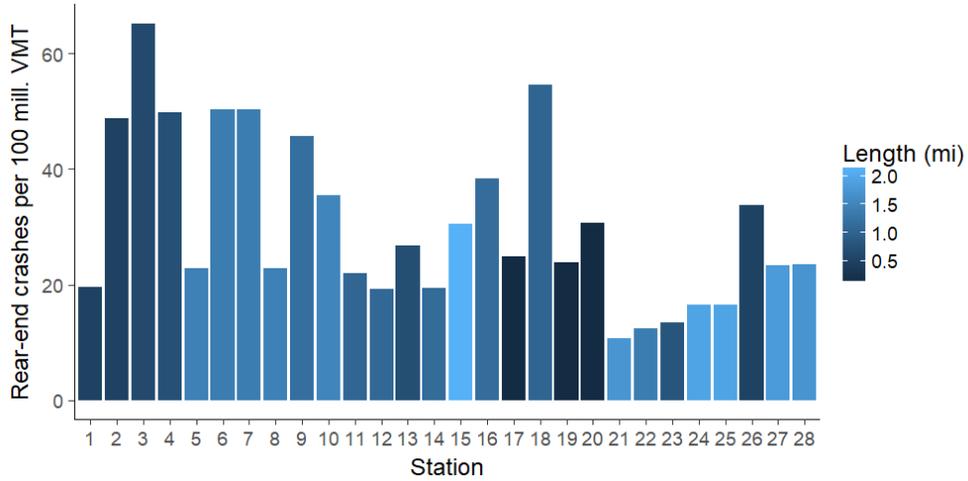

Figure 4 Rear-end crash rates for different sensor stations and the length of the freeway segments.

***Crash Rates vs. Drop in Required Reaction Time***

**Figure 5**(a) and (b) show the scatterplots of rear-end crash rates and the two drops in required reaction time– $\Delta t_2$ and $\Delta t_1$, respectively, as demonstrated in Figure 1. In order to investigate their correlation, three regression models were fitted to the data: an exponential model, linear model, and logarithmic model. The model forms are described in **Table 2**.

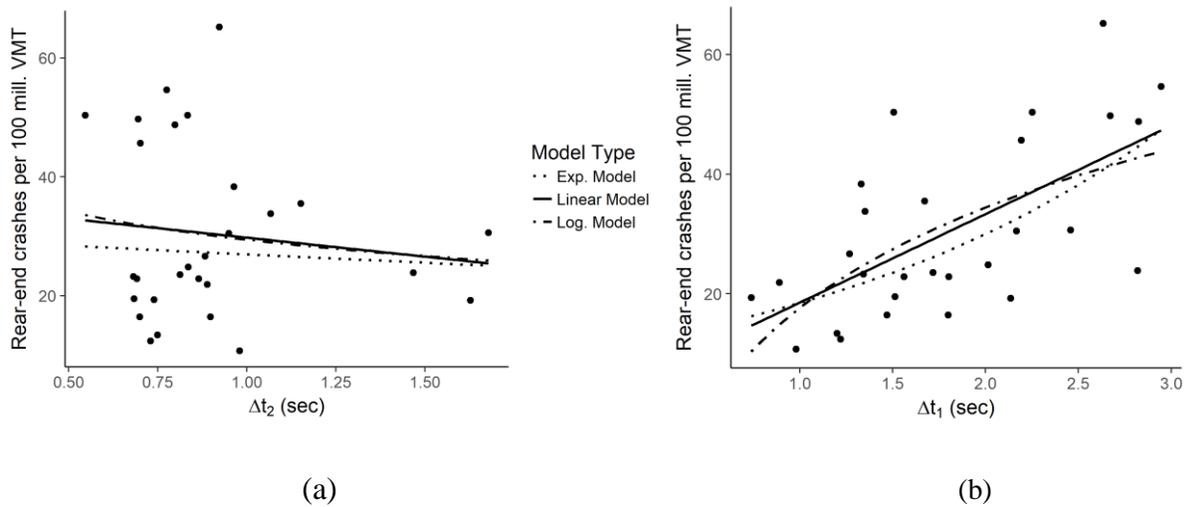

(a)                                                    (b)

Figure 5 Rear-end crash rates vs. (a) $\Delta t_2$ (b) $\Delta t_1$

It is apparent from **Figure 5**(a) that no correlation exists between the rear-end crash rates and $\Delta t_2$. It could be due to the fact that $\Delta t_2$ is very sensitive to the selection of break point density for the uncongested regime, which makes this predictor less robust to any potential outliers. All three fitted models demonstrated a very poor fit to the data, as the R-squares were found between 0.004 and 0.01.

On the other hand, a strong positive trend is apparent in the rear-end crash rates vs. $\Delta t_1$ plot as shown in **Figure 5**(b). The correlation coefficient between these two variables are +0.63. Since this evidence is mostly empirical, it is difficult to pre-determine about what should best explain the relationship. Nonetheless, all three forms of model showed a reasonable fit. **Table 2** shows the form of the fitted models, the fitted parameter values along with several statistical measures to evaluate the relationship.

TABLE 2 Regression Model Forms and Evaluations

| Model type | Model form | Residual SE | Adj. R-sq. | F-stat | Param. | Estimate | Std. Error | p-value |
|---|---|---|---|---|---|---|---|---|
| Exponential | $y = \mathrm{e}^{mx+c}$ | 11.63 | 0.37 | 17.04 | m | 0.48 | 0.12 | $3x10^{-4}$ |
|  |  |  |  |  | c | 2.43 | 0.22 | $3x10^{-11}$ |
| Linear | $y = mx + c$ | 11.53 | 0.38 | 17.34 | m | 14.79 | 3.55 | $3x10^{-4}$ |
|  |  |  |  |  | c | 3.74 | 6.77 | 0.59 |
| Logarithmic | $y = mln(x) + c$ | 11.86 | 0.34 | 14.98 | m | 24.20 | 6.25 | $6x10^{-4}$ |
|  |  |  |  |  | c | 17.63 | 3.99 | $2x10^{-4}$ |

Here, the response variable (y) is the rear-end crash rate (in crashes per 100 mill. VMT), the explanatory variable (x) is $\Delta t_1$ (in seconds), and m and c are model coefficients. The exponential and linear model exhibited similar performance measures in terms of the residual standard error (SE), adjusted R-square, and F-statistics. On the other hand, the logarithmic model, while showing a satisfactory fit, generated a slightly higher residual error and a lower R-square and F-statistic. The p-value column shows that except for the Y-intercept of the linear model, other coefficients of all three models are statistically significant at least at a level of 0.001.

Here, it is essential to explain the interpretation of the model evaluations from the perspective of this study. For instance, the R-square of 0.38 generated by the linear model interprets that the model explains 38% of the variation in the rear-end crash rates. Note that while $\Delta t_1$ is associated only with the transition from uncongested to the congested regime, not all rear-end crash rates happened during the transition regime. Moreover, a lot of such crashes are not directly related to the driver reaction time required to maintain asymptotic stability. Rather, several other factors including distracted driving, driving under the influence, animal crossing, and abrupt lane changing maneuver may lead to rear-end crashes, which are not reflected in the long-term traffic stream characteristics of a segment. Nonetheless, a lot of rear-end crashes are associated with the onset of congestion and the transition between the two regimes. Hence, only a significant portion of all rear-end crashes are expected to be explained by the fitted models. Having said that, an R-square value within the range of 0.34 to 0.38 appears to be satisfactory for the purpose of this study.

The variance of the residuals generated by a statistical model is another important criterion to evaluate statistical models. Despite having a satisfactory R-square, a model might be heteroscedastic if the residual shows a trend with the fitted response variable. To check for heteroscedasticity, the square root of standardized residuals vs. the fitted response variable plot, also known as the scale-location plot, is commonly used (University of Virginia Library, Charlottesville ). **Figure 6** shows such plots for all three fitted models along with a smoothed trend line and the confidence interval.

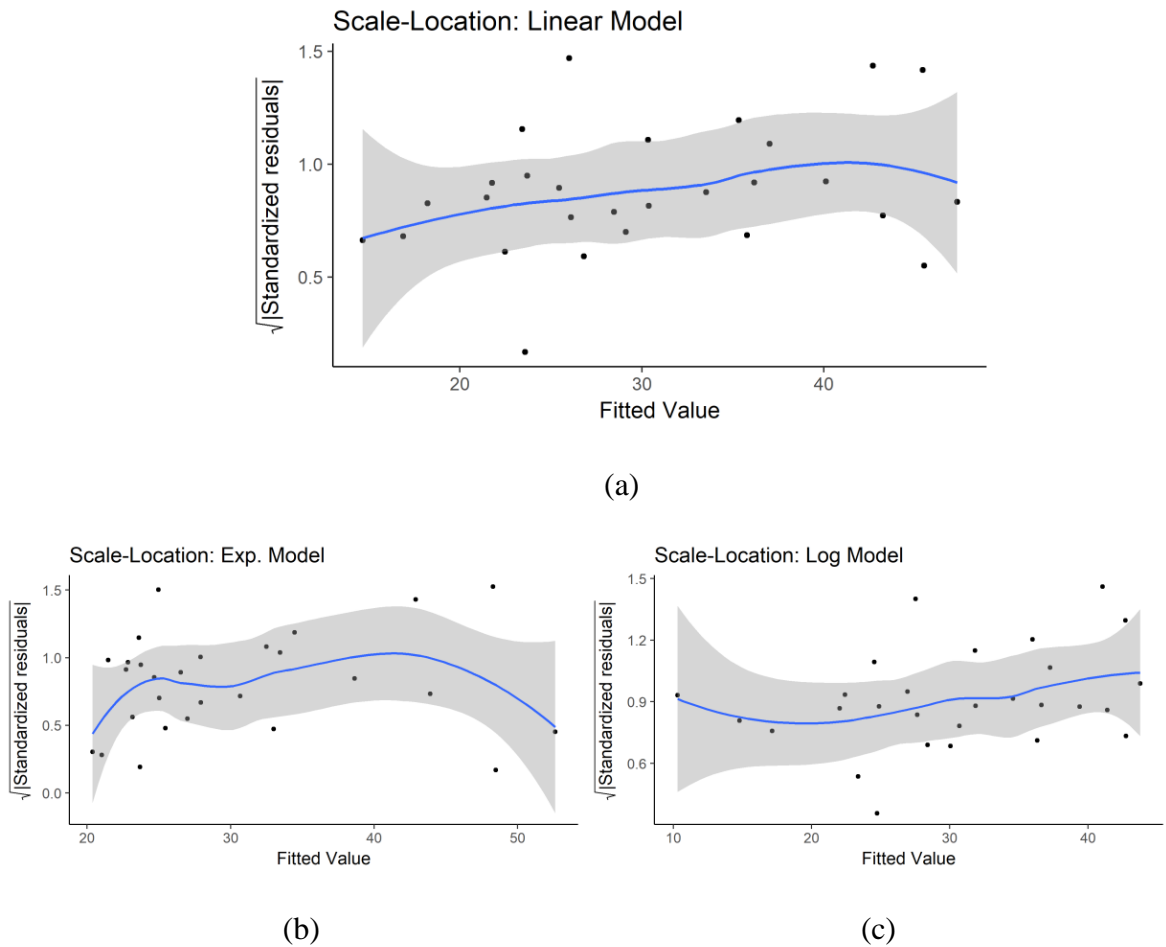

(a)

(b)                                             (c)

Figure 6 Scale-location plot for (a) Linear (b) Exponential and (c) Logarithmic model

**Figure 6** shows that the linear and the logarithmic model have a very mild increasing trend with the increase of the fitted values. However, the residuals are more uniformly distributed across the fitted values in the case of the linear model than the logarithmic model.

The scale-location plot for the exponential model has a stronger trend than the other two models which infers that the model has heteroscedasticity.

It is apparent that the linear model performed best among the three models both in terms of R-square value and homoscedasticity. It also has the simplest model form. The slope of the model interprets that for each second increment in $\Delta t_1$, the estimated rear-end crash rate increases by 14.79 crashes per 100 million VMT. Note that the high p-value of the intercept term of the linear model infers that the null hypothesis that the intercept is zero cannot be rejected. However, the intercept term is a trait obtained by extrapolating the linear model; the drop in driver reaction time cannot be zero in reality.

**CONCLUSIONS**

The study investigates the relationship between rear-end crash rates and macroscopically derived driver reaction time required for asymptotic stability. Separate GHR models are fitted discontinuously for the uncongested and congested flow regimes using traffic sensor data collected for one year. The drops in reaction time at the beginning and end of the transition regime are mainly focused here. Data from 28 sensors located in basic freeway segments of three interstates are used in this study. A freeway segment surrounding each sensor is selected for crash data analysis such that the road geometric characteristics remain constant throughout the segment. Rear-end crash rate for each segment is then estimated by examining police-reported crash data.

Results from the fitted models and crash data analysis showed that the selected sites have significant variability in terms of both traffic characteristics and rear-end crash risks. Standard deviations of the estimates were of small magnitude for the parameters that have a physical interpretation and to which the model is sensitive. Three model forms, namely linear, exponential, and logarithmic were used to test the correlation between crash rates and the drops in required reaction time. The second drop in required reaction time exhibited no significant correlation with rear-end crash rates. The first drop in required reaction time exhibited a strong positive correlation with rear-end crash rates. The correlation coefficient was found +0.63. The three models exhibited R-square values ranging from 0.34 to 0.38, with the linear model performing best both in terms of R-square and standard error. Analysis of the residuals for each model revealed that the linear and logarithmic model have reasonable homoscedasticity.

While an R-square within this range might seem low, it should be noted that not all rear-end crashes are associated with the transition of traffic regimes and car-following behavior. Hence, all rear-end crash occurrences at a location cannot be explained by the drop in required reaction time. Having said that, the statistical correlation revealed here can be considered strong enough to motivate follow-on research to incorporate macroscopically derived reaction time in safety and planning level studies.

More generally, this study demonstrates a useful application of a discontinuous macroscopic traffic model. However, a few limitations need to be highlighted here to be addressed in future research. A sample size of 28 falls slightly short of the generally accepted sample size of 30 for assumption of normality. As much as it is difficult to extract accurate crash data, it is important to increase the number of study sites. Further, if detailed data are available, it is recommended to filter the crashes that are directly associated with the transition of traffic regimes and car-following behavior.

**ACKNOWLEDGMENTS**

The authors would like to thank the National Transportation Center at Maryland for their financial support in this research

**AUTHOR CONTRIBUTIONS**

The authors confirm contribution to the paper as follows: study conception and design: B. Williams, I. Ahmed; data collection: I. Ahmed, S. Samandar, G. Chun; analysis and interpretation of results: I. Ahmed, B. Williams., S. Samandar; draft manuscript preparation: I. Ahmed, B. Williams, S. Samandar, G. Chun. All authors reviewed the results and approved the final version of the manuscript.

**REFERENCES**

Abdel-Aty, Mohamed, Nizam Uddin, Anurag Pande, M. Fathy Abdalla, and Liang Hsia. 2004. "Predicting Freeway Crashes from Loop Detector Data by Matched Case-Control Logistic Regression." *Transportation Research Record* 1897 (1): 88-95.

Ahmed, Ishtiak, Nagui M. Rouphail, and Shams Tanvir. 2018. "Characteristics and Temporal Stability of Recurring Bottlenecks." *Transportation Research Record* 2672 (42): 235-246.

Ahmed, Ishtiak, Billy M. Williams, and M. Shoaib Samandar. 2018. "Application of a Discontinuous Form of Macroscopic Gazis–Herman–Rothery Model to Steady-State Freeway Traffic Stream Observations." *Transportation Research Record* 2672 (20): 51-62.

Ahn, Soyoung, Jorge Laval, and Michael J. Cassidy. 2010. "Effects of Merging and Diverging on Freeway Traffic Oscillations: Theory and Observation." *Transportation Research Record* 2188 (1): 1-8.

American Association of State Highway and Transportation Officials. 2010. *Highway Safety Manual*. Vol. 1 AASHTO.

Brill, Edward A. 1972. "A Car-Following Model Relating Reaction Times and Temporal Headways to Accident Frequency." *Transportation Science* 6 (4): 343-353.

Chandler, Robert E., Robert Herman, and Elliott W. Montroll. 1958. "Traffic Dynamics: Studies in Car Following." *Operations Research* 6 (2): 165-184.

Chatterjee, Indrajit and Gary A. Davis. 2016. "Analysis of Rear-End Events on Congested Freeways by using Video-Recorded Shock Waves." *Transportation Research Record* 2583 (1): 110-118.

Davis, Gary A. and Tait Swenson. 2006. "Collective Responsibility for Freeway Rear-Ending Accidents?: An Application of Probabilistic Causal Models." *Accident Analysis & Prevention* 38 (4): 728-736.

Edie, Leslie C. 1961. "Car-Following and Steady-State Theory for Noncongested Traffic." *Operations Research* 9 (1): 66-76.

Gazis, Denos C., Robert Herman, and Richard W. Rothery. 1961. "Nonlinear Follow-the-Leader Models of Traffic Flow." *Operations Research* 9 (4): 545-567.

Kim, Taewan and H. Michael Zhang. 2011. "Interrelations of Reaction Time, Driver Sensitivity, and Time Headway in Congested Traffic." *Transportation Research Record* 2249 (1): 52-61.

Lord, Dominique, Abdelaziz Manar, and Anna Vizioli. 2005. "Modeling Crash-Flow-Density and Crash-Flow-V/C Ratio Relationships for Rural and Urban Freeway Segments." *Accident Analysis & Prevention* 37 (1): 185-199.

May, Adolf D. 1965. "Traffic Flow Theory-the Traffic Engineers Challenge." *Proc.Inst.Traf.Eng*: 290-303.

Misener, James A., H-S Jacob Tsao, Bongsob Song, and Aaron Steinfeld. 2000. "Emergence of a Cognitive Car-Following Driver Model: Application to Rear-End Crashes with a Stopped Lead Vehicle." *Transportation Research Record* 1724 (1): 29-38.

North Carolina Department of Transportation. "Traffic Engineering Accident Analysis System (TEAAS).", accessed 4/21/, 2020, https://connect.ncdot.gov/resources/safety/Pages/TEAAS-Crash-Data-System.aspx.

Oh, Cheol and Taejin Kim. 2010. "Estimation of Rear-End Crash Potential using Vehicle Trajectory Data." *Accident Analysis & Prevention* 42 (6): 1888-1893.

Pande, Anurag and Mohamed Abdel-Aty. 2006. "Comprehensive Analysis of the Relationship between Real-Time Traffic Surveillance Data and Rear-End Crashes on Freeways." *Transportation Research Record* 1953 (1): 31-40.

Smith, Ralph C. 2013. *Uncertainty Quantification: Theory, Implementation, and Applications*. Vol. 12 Siam.

Tanaka, Mitsuru, Prakash Ranjitkar, and Takashi Nakatsuji. 2008. "Asymptotic Stability and Vehicle Safety in Dynamic Car-Following Platoon." *Transportation Research Record* 2088 (1): 198-207.

Tanvir, Shams, R. T. Chase, and N. M. Roupahil. 2019. "Development and Analysis of Eco-Driving Metrics for Naturalistic Instrumented Vehicles." *Journal of Intelligent Transportation Systems*: 1-14.

Touran, Ali, Mark A. Brackstone, and Mike McDonald. 1999. "A Collision Model for Safety Evaluation of Autonomous Intelligent Cruise Control." *Accident Analysis & Prevention* 31 (5): 567-578.

Transportation Research Board of the National Academies. 2016. *Highway Capacity Manual, Sixth Edition: A Guide for Multimodal Mobility Analysis* . 6th ed. Washington, D.C.:.

Xu, Tu and Jorge A. Laval. 2019. "Analysis of a Two-Regime Stochastic Car-Following Model: Explaining Capacity Drop and Oscillation Instabilities." *Transportation Research Record* 2673 (10): 610-619.

Zhang, H. Michael and T. Kim. 2005. "A Car-Following Theory for Multiphase Vehicular Traffic Flow." *Transportation Research Part B: Methodological* 39 (5): 385-399.

Zielke, Benjamin A., Robert L. Bertini, and Martin Treiber. 2008. "Empirical Measurement of Freeway Oscillation Characteristics: An International Comparison." *Transportation Research Record* 2088 (1): 57-67.